# 'Being there together for health': A Systematic Review on the Feasibility, Effectiveness and Design Considerations of Immersive Collaborative Virtual Environments in Health Applications

Tohid Zarei[1]*, Michelle Emery[1], Dimitrios Saredakis[1], Gun A. Lee[2]^, Ben Stubbs[3]^, Ancret Szpak[1,4]^, Tobias Loetscher[1]

1 Cognitive Ageing and Impairment Neurosciences (CAIN) Laboratory, Behaviour-Brain-Body (BBB) Research Centre, UniSA Justice & Society, University of South Australia (UniSA), SA, Australia

2 Empathic Computing Lab, IVE: Australian Research Centre for Interactive and Virtual Environments, UniSA STEM, University of South Australia (UniSA), SA, Australia

3 UniSA Creative, University of South Australia (UniSA), SA, Australia

4 School of Psychology, Faculty of Health and Medical Sciences, The University of Adelaide (UoA), SA, Australia

*Correspondence to: tohid.zarei@mymail.unisa.edu.au; University of South Australia Magill Campus, St Bernards Road, Magill, SA 5072, Australia

^Equal Contribution

## Abstract

Effectively using immersive multi-user environments for digital applications (via virtual, augmented and mixed reality technologies) beckons the future of healthcare delivery in the metaverse. We aimed to evaluate the feasibility and effectiveness of these environments used in health applications, while identifying their design features.

We systematically searched MEDLINE, PsycINFO, and Emcare databases for peer-reviewed original reports, published in English, without date restrictions until Aug 30, 2023, and conducted manual citation searching in Feb 2024. All studies using fully immersive extended reality technologies (e.g., head-mounted displays, smart glasses) while engaging more than one participant in an intervention with direct health benefits were included. A qualitative synthesis of findings is reported. The quality of research was assessed using JBI Critical Checklists. The review was pre-registered on PROSPERO (CRD42023479155).

Of 2862 identified records, 10 studies were eligible, with high average quality ratings of quantitative (85%) and qualitative (74%) data analyses. Included studies were mostly conducted with healthy young adults (five studies) and older adults (four studies). They were aimed at various well-being promotion, symptom reduction, and skill acquisition applications. While they all used different models of Oculus/Meta headsets, their environments' designs were distinctive and aligned with their objectives. Findings indicated varying degrees of positive health outcomes, for engagement in rehabilitation, meaningful interactions across distances, positive affect, transformative experiences, mental health therapies, and motor skill learning. Participants reported high usability, motivation, enjoyment, presence and copresence. They also expressed the need for more training time with technology.

It is difficult to make definitive assertions at this stage about the effectiveness of immersive shared virtual environments in health programs due to heterogeneous methodologies and aims. However, adopting an intentional intervention design, considering factors affecting presence and copresence, as well as integrating co-creation of the program with participants, seems integral to achieving positive health outcomes.

## Introduction

Immersive technologies are gaining momentum in the health sector, with applications becoming increasingly widespread and sophisticated (Loetscher et al., 2023; Philippe et al., 2022). These technologies are collectively known as extended reality (XR), which include virtual reality (VR), augmented reality (AR) and mixed reality (MR). While there is a relatively established body of evidence for effectively using single-user environments in health interventions (Bansal et al., 2022), research on shared XR environments—where multiple users interact simultaneously—is limited. This is in spite of the fact that the enabled human interactions in these collaborative virtual environments (CVEs) open a huge window of opportunity for engaging and effective applications (Steed & Schroeder, 2015)[1].

---

[1] Unless otherwise specified, the terms 'shared virtual environment (SVE)', 'collaborative virtual environment (CVE)', and 'multi-user virtual environment (MUVE)' are used interchangeably throughout this text.

One contributing factor to the relatively fewer and more crude instances of CVEs in the literature might be the limited technological affordances of previous non-immersive desktop-based systems (Weiss & Klinger, 2009). With the widespread availability of head-mounted displays (HMDs), smart glasses and other XR headsets, which provide users with a fully immersive experience, there now exists the possibility of heading towards multi-user XR health interventions.

Tapping into this potential may revolutionise healthcare delivery, especially with the advent of the 'metaverse' (Thomason, 2021), defined as 'a 3D transcendent world that converges the physical, digital, and biological worlds using a hyper-connected, intelligent new technology that impacts every economy and industry' (Lee, 2022, p. 2). Immersive technologies are essential to metaverse services, combining virtuality and reality. Indeed, 'Big Tech' companies have found metaverse as one of their 'long-term bets on the technologies of the future' and invested accordingly and quite competitively on it (Apple, 2023; Bosworth, 2023; Shaw, 2021). The global metaverse market is projected to reach a value of US$507.8 billion by 2030, jumping with an annual growth rate of 37.73% from US$74.4 billion in 2024 (Statista, 2024).

The term 'virtual reality' was coined by Jaron Lanier (Barlow et al., 1990); the founder of VPL Research company which was among the pioneers in developing VR products—including perhaps the very first immersive collaborative virtual environment (ICVE) via HMDs (Blanchard et al., 1990; Slater & Sanchez-Vives, 2016). Twenty years earlier, the prototype of an immersive VR system had been already introduced and described by computer scientist Ivan E. Sutherland, by surrounding users with two-dimensional images to create the illusion of seeing 3D objects (Sutherland, 1968). While they utilised a quite different technological setup, the essential conditions for providing immersion to users have remained similar to this day. Research has shown an array of factors determining immersion, which is the function of a VR system's properties—including, most importantly, the use of head tracking, and stereoscopic display with a wide field of view (Cummings & Bailenson, 2016).

However, immersion does not fully capture how participants experience the virtual environment (VE) into which they are catapulted. As the subjective correlate of immersion, presence is a term referring to the psychological state of participants in a VE; simply put, an illusion of 'being there' (Slater, 2009). It was shown that presence is facilitated when authentic behavioural, emotional and cognitive responses, and sense of agency are evoked in participants while engaging with the environment (Riches et al., 2019). Another relevant concept, specific to VEs with more than one participant, is copresence: the feeling of being there with others (Schroeder, 2006). Three factors were mentioned by Steed and Schroeder (2015) as affecting copresence: modality (i.e., substitution of different sensory modalities, mainly visual and auditory), realism (i.e., behavioural realism of users' representation in the CVE, their avatars, in interaction with others, realised through naturalistic eye direction, facial expressions and gestures), and context (i.e., familiarity of people to each other, task of the CVE, and size of the group).

Note that the degree of fidelity and photorealism in VEs, in and by itself, does not determine levels of presence and copresence. In fact, even basic VEs from the early 1990s, despite their very limited photorealism, were capable of eliciting high levels of presence (Conn et al., 1989). This is particularly relevant in clinical and healthcare applications of immersive technologies (Bouchard & Rizzo, 2019), so as not to equate the use of advanced technology with the effectiveness of the programs. Nonetheless, considering factors that influence presence and copresence, as well as customising them to align with the specific goals of a program and its target population can result in more refined ICVEs for health interventions. There is a delicate balance that needs to be established between providing an engaging experience for users and ensuring the effectiveness of the intervention.

Applications of XR technologies in various health domains stretch from surgical operations and medical training to rehabilitation and mental health (Lee, 2022; Loetscher et al., 2023). Accordingly, there are already comprehensive as well as domain-specific reviews and surveys in the literature (e.g., Bansal et al., 2022; Philippe et al., 2022). However, the focus of this review is on applications that utilise these technologies to actively engage more than one participant in an ICVE-based program with direct health benefits for users. Considering single-user immersive virtual environments (IVEs), there have been promising results for diverse health applications. In mental health disorders, including phobia (Eshuis et al., 2021), post-traumatic stress disorder and anxiety disorders (van Loenen et al., 2022), IVE-based interventions were shown to be more effective than control conditions and comparable to standard interventions. They were also used effectively to manage pain during medical procedures, mainly via distraction (Indovina et al., 2018). Furthermore, there is documented success in utilising IVEs for the rehabilitation of various physical and cognitive conditions, such as post-stroke rehabilitation (Demeco et al., 2023; Rose et al., 2018).

Given that ICVEs introduce unique complications and require careful consideration across various applications, designs, and health aims, conducting an up-to-date review is imperative. Consequently, the aim of this systematic review is to summarise the body of evidence and provide a qualitative synthesis of studies that utilised ICVEs, using XR technologies, to provide health benefits to participants. To this end, the following objectives are realised by this systematic review:

1)      To evaluate the reported applications, feasibility and/or effectiveness outcomes of ICVEs in health-related (including physical and mental health, and rehabilitation) interventions and programs with typical or clinical populations.

2)      To identify the design characteristics of the ICVEs in such studies.

## Methods

A systematic literature search was conducted in accordance with Preferred Reporting Items for Systematic Reviews and Meta-Analysis (PRISMA) guidelines (Moher et al., 2015; Page et al., 2021). The review protocol was pre-registered in PROSPERO (ID: CRD42023479155; available from: https://www.crd.york.ac.uk/prospero/display_record.php?ID=CRD42023479155). Three electronic databases were searched on the 30th August 2023: MEDLINE (via the Ovid interface), APA PsycINFO (via the Ovid interface), and Emcare (via the Ovid interface). The search terms included various combinations of keywords related to XR technologies (e.g., "virtual reality", "augmented reality", and "head-mounted display"), and their collaborative aspects (e.g., "collaborative", "multi-user", and "shared virtual environment"). The search was adapted to each database's specific subject headings and syntax requirements. No filters or limitations were applied to the search engine interface. The complete search string is accessible through the review's PROSPERO record. To capture studies that might not have been identified through the initial database searches, hand searches of relevant reviews' bibliographies and authors' personal files were conducted, as well as backward and forward citation searching of the included studies in January and February 2024.

### Eligibility Criteria

The inclusion and exclusion criteria were drafted and refined by three review authors (TZ, TL, and DS). Studies were considered eligible by meeting the following criteria:

*Population:*

1) All clinical or non-clinical populations in all age ranges have been considered.

*Intervention*:

2) An IVE must have been used as part of the intervention (via 3D XR technologies, including VR, AR and MR technologies, for example, HMDs, smart glasses and other wearables, cave automatic virtual environments). Accordingly, studies solely using non-immersive displays (such as desktop computers with two-dimensional displays) were excluded.
3) The IVE intervention, program or training must have been aimed at having direct health benefits for participants by engaging them in the program. Studies utilising IVE only for assessment, therefore not aimed at conducting any interventions, were not considered. Neither were studies aimed at training health professionals.
4) The IVE must have been collaborative or multiuser, meaning at least two persons have been immersed and interacted with one another synchronously at any time point inside the environment. The study was excluded if only one user at a time was immersed or there was no interaction of any sort between users while immersed.

*Comparator*:

5) All comparisons have been included (e.g., pre-post, waitlist, treatment-as-usual [TAU], active control group, qualitative analysis).

*Outcome*:

6) Health-related (e.g., symptom reduction, well-being promotion, skill acquisition) and/or feasibility (including practicality or user experience) outcomes must have been reported. In case of qualitative studies, the collected data and reported findings must have been related to either or both outcome domains mentioned.

*Study design:*

7) All research designs have been included (e.g., randomised controlled trials [RCTs], quantitative or qualitative pilot studies, case studies).

*Report characteristics:*

8) Reports must have been published in English or translated for publication.
9) All years of publication have been considered.
10) Reports must have been original and peer-reviewed with full-text available (including journal and conference papers but excluding reviews, dissertations, only abstracts).

## Selection & Data Extraction Process

A PRISMA-compatible flowchart is shown in Figure 1. Systematic search records corresponding to each database (along with articles collected through other sources mentioned above) were uploaded in separate files into Covidence systematic review software (Veritas Health Innovation, Melbourne, Australia; available at www.covidence.org). De-duplication of repeated references was done automatically by the software. In total, 2855 records were retrieved from database searching and 7 from other sources, from which 540 references were removed by Covidence's de-duplication tool.

Three authors (TL, DS, and TZ) contributed to the title-and-abstract screening stage while considering the eligibility criteria. The vote of at least two independent authors was necessary for a study to be identified as eligible at this stage. Full reports were obtained for 52 titles that met the criteria or if there was uncertainty in deciding based on their title and abstract. Conflicts in authors' votes were reviewed in meetings following each stage until a consensus was reached. The same process with a review author pair (TZ, ME, and TL) was carried out for the full-text screening stage, with the addition of recording reasons for excluding reports. Reviewers were not blind to the journal titles or the reports' authors or institutions.

Two independent reviewers (TZ and ME) used Covidence software to extract data from the 10 included studies using a pre-planned form based on agreed-upon data items. Discrepancies were flagged, discussed, and resolved in subsequent meetings. Missing data items were requested from corresponding authors via email, with two attempts made before marking them as missing.

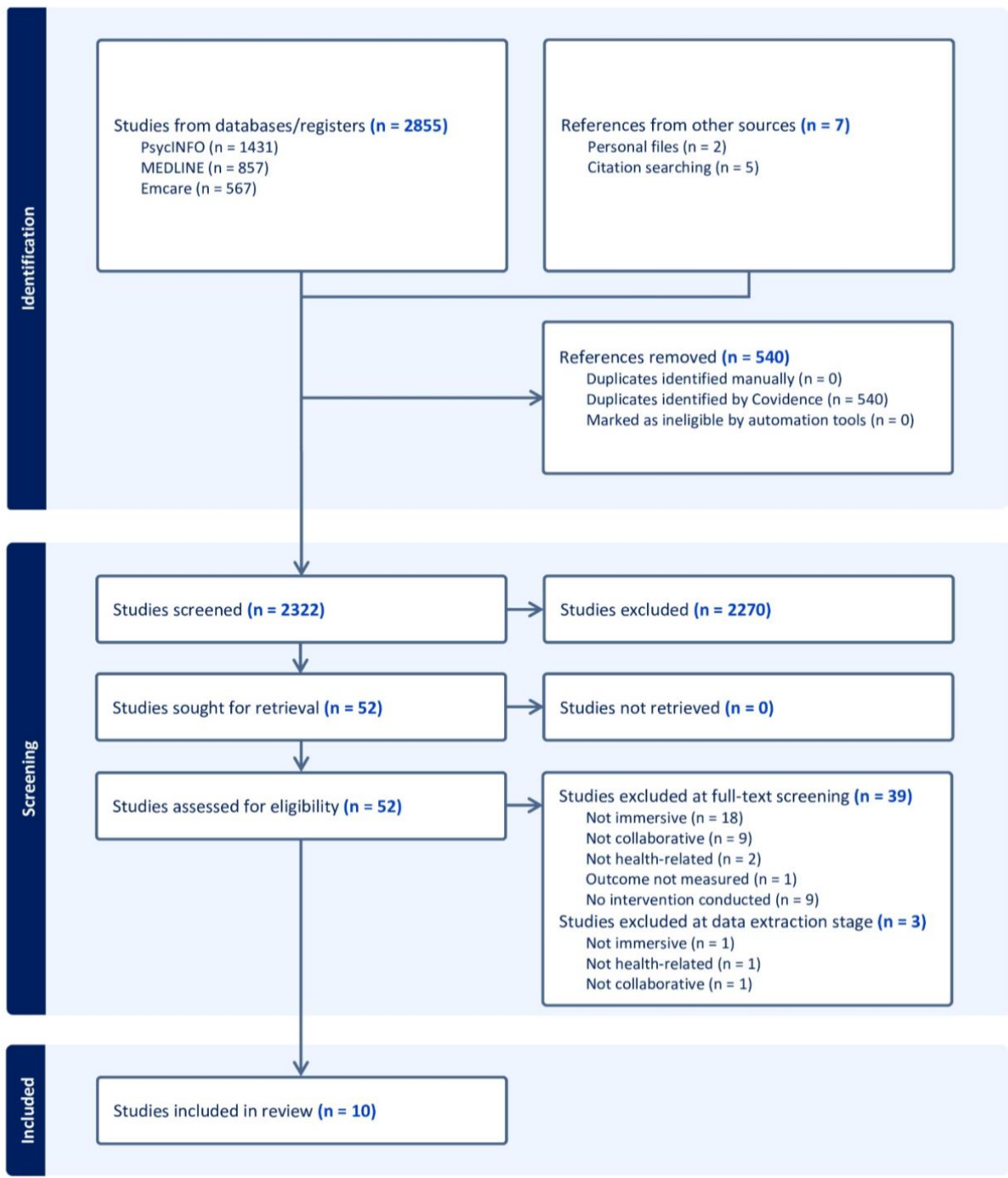

*Figure 1. PRISMA flowchart of the systematic review.*

## Data & Outcome Items

The following items were extracted from the included studies:

Data items: 1) *Study and target population characteristics:* country, age range (mean), total sample size (% gender), diagnosis or characteristic feature (if any), study design; 2) *Intervention (i.e., both ICVE and comparator/s):* name, technology used (a summary description of the hardware and software configurations), physical setting (e.g., clinic, school), physical distance of participants (i.e., remote, or co-located), type of health aim/s (e.g., symptom reduction, well-being promotion, skill acquisition), health outcome variable scales (if any), feasibility outcome variable scales (if any), number and duration of session/s, duration of follow-up (if any), cybersickness (i.e., adverse and negative effects of the technology).

Outcomes: 1) *Results of effectiveness and/or feasibility (e.g., practicality, acceptability, usability, evaluations) measures of the ICVE intervention or program, and the comparator (if any)*; 2) *Design characteristics of the ICVE*: virtual setting (e.g., a virtual replica of a cafe), number of simultaneous users, modes of interaction in the ICVE (e.g., verbal, shared endeavour), task or objective of the environment, support provided (e.g., diagnosis-specific, presence of facilitators), design factors used and their instantiation in the environment, based on Dalgarno and Lee (2010), and Glaser and Schmidt (2021): representational fidelity (i.e., realistic display of environment, smooth display of view changes and object motion, consistency of object behaviour, user representation, spatial audio, kinaesthetic and tactile force feedback), and user interaction (i.e., embodied actions, embodied verbal and non-verbal communication, control of environment attributes and behaviour, construction/scripting of objects and behaviours).

## Data Synthesis

Given the heterogeneity of studies in this emerging field, and the focus of the review on design considerations, qualitative synthesis and summary, primarily based on the two objectives of this review, have been reported for the included studies.

## Quality Assessment

While, as outlined in the protocol, our initial intention was to utilise a single quality assessment form for all included studies, specifically the JBI Critical Appraisal Checklist for Quasi-Experimental Studies (Aromataris & Munn, 2020), we encountered challenges during the full-text screening and data extraction phases. Many items within this form were found to be non-applicable to studies predominantly reporting qualitative findings. Consequently, we decided to incorporate an additional assessment tool, namely the JBI Critical Appraisal Checklist for Qualitative Research (Aromataris & Munn, 2020), to evaluate the qualitative data analyses of the included studies.

As a result, depending on whether a study reported quantitative, qualitative, or both types of findings, either or both quality ratings were assigned accordingly.

The JBI Critical Appraisal Checklist for Quasi-Experimental Studies and the JBI Critical Appraisal Checklist for Qualitative Research consist of 9 and 10 items, respectively, each with four response options (Yes/No/Unclear/Not Applicable). In cases where an item was deemed not applicable to the study design of a particular report, it was not included in the total rating. An 'Unclear' response was treated as equivalent to a 'No' response.

A total score for each quality rating was calculated by assigning a score of 1 for each 'Yes' response and 0 for each 'No' response in the quality assessment. The obtained summed score was then divided by the total possible score on the respective checklist, resulting in a summary score between 0 and 1. By adopting this approach, reports were evaluated based on their specific study designs.

TZ and ME assessed and independently rated each study accordingly. Any discrepancies were resolved in subsequent meetings.

# Results

## Study Characteristics and Target Populations

Table 1 summarises the information regarding characteristics and target populations of the reviewed studies. Five of the ten included studies (50%) adopted a non-randomised experimental research design (Barberia et al., 2018; Guertin-Lahoud et al., 2023; Høeg et al., 2023; Kalantari et al., 2023; Robinson et al., 2023), all of which, except for one (Barberia et al., 2018) which only reported on quantitative findings, documented both quantitative and qualitative data collected through tasks, questionnaires and interviews. Two studies (20%) were randomised controlled trials (Kodama et al., 2023; Shah et al., 2023). One of these studies reported both quantitative and qualitative data (Shah et al., 2023). The remaining three studies (30%) employed qualitative research methods (Baker et al., 2021; Li & Yip, 2023; Matsangidou et al., 2022), focusing primarily on interview findings. Among them, one was a case study (Li & Yip, 2023). Studies were conducted in various countries: three in Europe (Norway, Spain, and the UK [jointly with Cyprus]), three in North America (Canada and the US), two in Asia (China and Japan), and two in Australia. Article publication years ranged from 2018 to 2023, with seven out of ten studies published in 2023.

*Table 1. Study characteristics and target populations.*

| Study | Country | Name of Program | Aim | Study Design | Target Population | Total No. of Participants (age mean/range; %female) |
|---|---|---|---|---|---|---|
| Baker et al. (2021) | Australia | School Days | The use of a bespoke social VR to support group reminiscence | Qualitative research | Older adults (over the age of 70) | 16 (70-81; 45%) |
| Høeg et al. (2023) | Australia | Buddy Biking | A virtual rehabilitation experience aimed to affect the users' motivation, engagement and performance | Non-randomised experimental study | Current or discharged patients (older adults) from a public rehabilitation outpatient service | 11 (60; 36%) |
| Robinson et al. (2023) | United States | Cognitive Behavioral Immersion (CBI) | To incorporate cognitive behavioural tools into an immersive massive multiplayer online (MMO) application | Non-randomised experimental study | Innerworld platform users (aged over 21) self-reported as being in recovery from a substance use disorder (SUD) | 48 (>21; 22%) |
| Li & Yip (2023) | Hong Kong, China | Remote Arts Therapy | A custom-designed collaborative virtual environment (CVE) to enable remote arts therapy | Case study | Early- to mid-20s adults with a moderate or high level of stress | 3 (early to mid 20s; 67%) |
| Guertin-Lahoud et al. (2023) | Canada | International Space Station (ISS) | To explore the perceived and lived experience of users, individually or in dyads, in a VR experience comprising different levels of interactivity | Non-randomised experimental study | Visitors of a multimedia entertainment centre | 28 (24.7; 43%) |

Systematic Review on Immersive Collaborative Virtual Environments in Health

| Kalantari et al. (2023) | United States | Social-VR Program | To enable VR-mediated interactions to produce meaningful social experiences | Non-randomised experimental study | Older adults (over the age of 60) | 36 (71; 79%, 3% Nonbinary) |
|---|---|---|---|---|---|---|
| Shah et al. (2023) | Norway | Social VR-based Exergame | Motivating elderly individuals to participate in physical exercise and improving social connectedness during rehabilitation | Randomised controlled trial | Past or present clients of a municipal rehabilitation centre (over the age of 60) | 14 (77.6; 50%) |
| Kodama et al. (2023) | Japan | Virtual Co-embodiment | To test whether the virtual co-embodiment improves motor skill learning efficiency | Randomised controlled trial | Healthy adults | 64 (23.5; 17%) |
| Barberia et al. (2018) | Spain | The Island | To explore how the VR simulation impacts death anxiety and changed attitudes to life | Non-randomised experimental study | Female Psychology undergraduate students speaking Catalan as mother tongue | 32 (20.1; 100%) |
| Matsangidou et al. (2022) | United Kingdom and Cyprus | MUVR Remote Psychotherapy | Using Multi-User Virtual Reality (MUVR) for remote psychotherapy | Qualitative research | Female undergraduate students at high-risk for developing an eating disorder | 14 (19.9; 100%) |

Regarding the populations targeted by the studies, four out of ten aimed to provide health benefits for older adults (see Figure 2), including enhancing motivation and engagement in physical rehabilitation (Høeg et al., 2023; Shah et al., 2023) and supporting meaningful social interactions (Baker et al., 2021; Kalantari et al., 2023). The remaining studies were predominantly aimed at healthy young adults (Kodama et al., 2023), with moderate to high stress levels (Li & Yip, 2023), in the real-life context of an entertainment centre (Guertin-Lahoud et al., 2023), and self-identified as being in recovery from a substance use disorder (Robinson et al., 2023). Two studies were conducted exclusively with female undergraduate students (Barberia et al., 2018; Matsangidou et al., 2022), including one with those at high-risk to develop an eating disorder (Matsangidou et al., 2022).

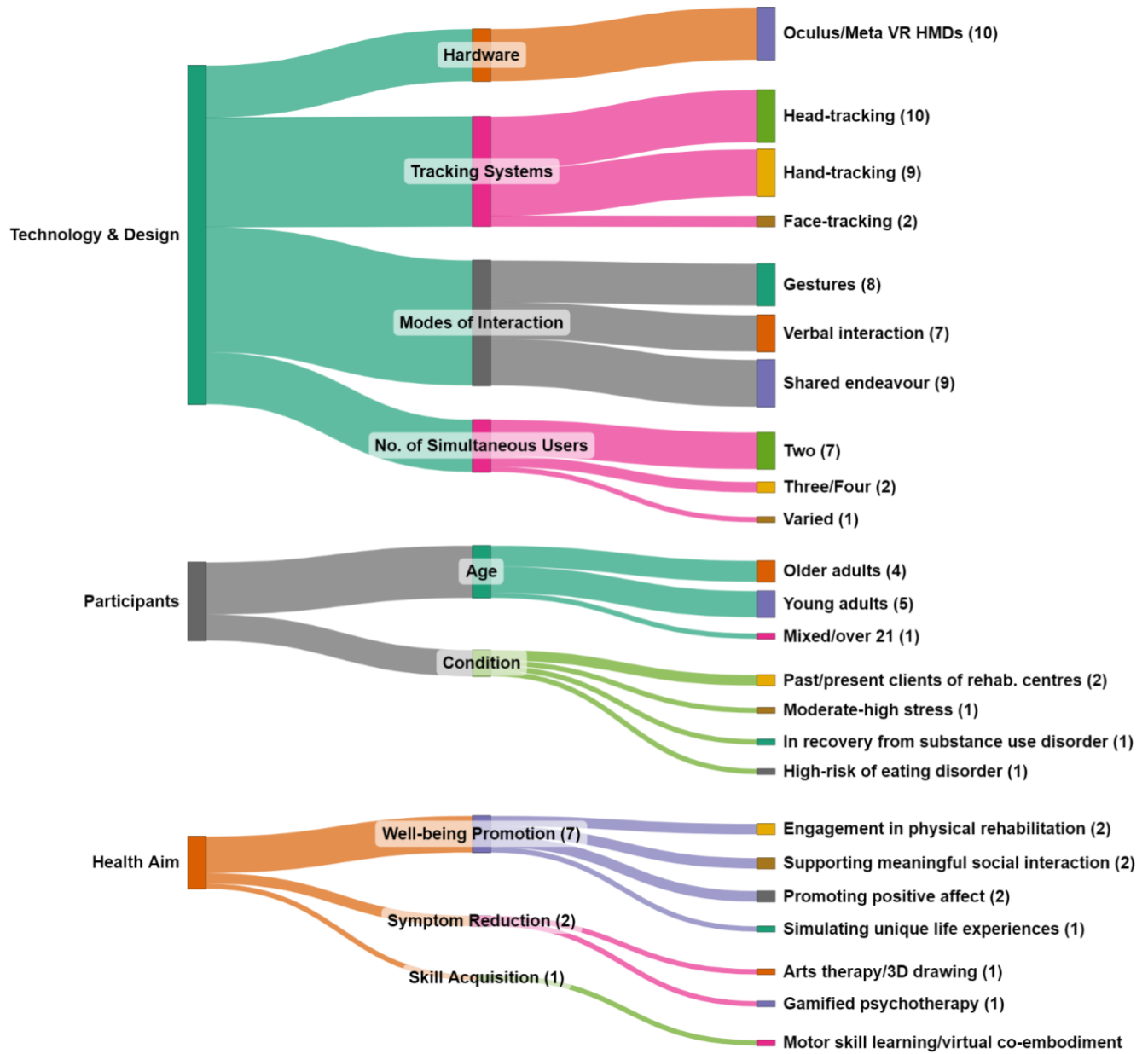

*Figure 2. Simplified Sankey diagram of the elements and factors in the technology-participant-health network based on the reviewed studies; Numbers in parentheses show the number of studies utilising or corresponding to that factor.*

**Interventions & Contexts**

Table 2 provides a description of the physical contexts of the ICVE interventions utilised in the reviewed studies.

*Hardware.* All of the studies used various models of Oculus HMDs (Meta Inc.; formerly Facebook) to provide fully immersive experiences for participants, ranging from the older Oculus Rift Development Kit 2 (Barberia et al., 2018) to the newer Oculus Quest 2 (Guertin-Lahoud et al., 2023; Kalantari et al., 2023; Kodama et al., 2023; Li & Yip, 2023). In addition to head-tracking which was incorporated in all ICVEs (by headsets' internal sensors), varying degrees of hand and body tracking were integrated as well. With the exception of Buddy Biking (Høeg et al., 2023), all programs mapped hand movements into the VEs, either with handheld controllers (Baker et al., 2021; Kalantari et al., 2023; Kodama et al., 2023; Li & Yip, 2023; Matsangidou et al., 2022; Robinson et al., 2023) or without them (Barberia et al., 2018; Guertin-Lahoud et al., 2023; Shah et al., 2023).

*Software.* With respect to the software utilised for creating the ICVEs, Unity (https://unity.com) was the most common option, used in seven of the studies to develop customised VEs (Baker et al., 2021; Barberia et al., 2018; Høeg et al., 2023; Kodama et al., 2023; Li & Yip, 2023; Matsangidou et al., 2022; Shah et al., 2023). The rest of the studies used an existing social VR platform (Spatial; Kalantari et al., 2023), a massive multiplayer online (MMO) application (Innerworld; Robinson et al., 2023), and a VR experience in an entertainment centre (Guertin-Lahoud et al., 2023), out of which, Kalantari and colleagues (2023) customised the VEs specific to their study. Notably, two studies (Baker et al., 2021; Matsangidou et al., 2022) integrated synchronous lip movements and verbal communication to animate the avatars' facial expression in their VEs.

*Physical context.* Five out of ten studies were conducted in research laboratories and offices (Baker et al., 2021; Kalantari et al., 2023; Kodama et al., 2023; Li & Yip, 2023; Matsangidou et al., 2022), while two took place in rehabilitation centres (Høeg et al., 2023; Shah et al., 2023), and one occurred in a multimedia entertainment centre (Guertin-Lahoud et al., 2023). The majority of studies adopted a remote physical setup, with participants simultaneously immersed in the SVEs but not in close physical proximity to each other (Baker et al., 2021; Barberia et al., 2018; Kalantari et al., 2023; Li & Yip, 2023; Matsangidou et al., 2022; Robinson et al., 2023; Shah et al., 2023). In the other three studies, participants were co-located in the same room (Guertin-Lahoud et al., 2023; Høeg et al., 2023; Kodama et al., 2023). The number and duration of ICVE intervention sessions varied across studies depending on their aims and research designs, with sessions ranging from one to ten and durations spanning 10 to 60 minutes per participant (see Table 2). On average, an ICVE session lasted approximately 37 minutes across studies, with four studies having sessions lasting around 45 minutes (Baker et al., 2021; Kalantari et al., 2023; Kodama et al., 2023; Li & Yip, 2023).

*Cybersickness.* Six studies screened for and/or reported potential adverse effects of HMDs on participants using standardised measures, interview findings, or by tracking dropouts (see Table 2; Barberia et al., 2018; Guertin-Lahoud et al., 2023; Høeg et al., 2023; Kalantari et al., 2023; Robinson et al., 2023; Shah et al., 2023). Notably, two participants across separate studies had to terminate their sessions prematurely due to discomfort and dizziness (Barberia et al., 2018; Høeg et al., 2023).

*Table 2. Physical contexts of the ICVE interventions.*

| Name of Program (Name of ICVE Intervention Group, if multiple groups) | Technology (hardware and software) | Physical Setting | Physical Distance of Participants | Number and Duration of Session/s | Reported Cybersickness |
|---|---|---|---|---|---|
| School Days (Baker et al.,2021) | - Oculus Rift headset connected to a gaming PC; two hand controllers; and two Rift room tracking sensors to map hand and torso movements.<br>- A bespoke VE built with Unity3D enabling up to four people to meet simultaneously; voice captured by a microphone and relayed to other users via speakers in the headset; facial and body (via the Final IK plugin by Root Motion) and lip movements (via the OVRLipSync plugin by Oculus) were integrated. | Research lab or office (at two different locations) | Remote (geographically distributed across two sites) | A total of 26 sessions, each of which lasted 27–56 minutes (average = 44 minutes). Each session involved two or three participants alongside a facilitator. | None reported |
| Buddy Biking (Høeg et al., 2023) | - Oculus Rift Consumer Version 1 (CV1) headset connected to a high-end gaming laptop; exercise bikes were both recumbent and regular upright exercise bicycles.<br>- VE was created using Unity3D and the EasyRoads3D asset tool to create virtual paths; system transformed the participant's cadence into forward movement in the VE; road friction, wind-whistling and chain ring clicking acoustics were incorporated. | Rehabilitation centre | Co-located | One session for each pair of participants (biked for 10.6 ± 2.6 mins) | One of the participants (P06) felt sick after a few minutes and had to quit biking. They experienced dizziness, stating that the hills and corners within the virtual environment gave a sensation of being on a roller coaster. |

| Cognitive Behavioral Immersion (Robinson et al., 2023) | - Oculus Quest headset and handheld controllers.<br>- CBI was delivered on a gamified metaverse application called 'Innerworld' (developed by Innerworld, Inc.); users communicated through a microphone on the headset. | Various (not specified) | Remote | Participants varied in the number of sessions they attended as well as the number of measures they completed. Each event in Innerworld lasted about 60 minutes. | Five participants experienced negative physiological impacts (e.g., nausea, headaches) associated with using VR for the first time. |
|---|---|---|---|---|---|
| Remote Arts Therapy (Li & Yip, 2023) | - Oculus Quest 2 headset with controllers; one desktop computer and one WiFi router.<br>- Software was developed using Unity; collaborative art creation interface was inspired by the design of Tilt Brush; interface followed the position and orientation of the controller in the participant's non-dominant hand; other controller could be used to draw or to interact with the user interface via its trigger buttons. | Laboratory (participants) and home (therapist) | Remote | Eight 45-minute one-on-one sessions with the therapist for eight consecutive weeks for each participant | None reported |
| International Space Station (Duo group; Guertin-Lahoud et al., 2023) | - Oculus Quest 2 headset; color-coded tracking system (body and hand tracking); audio and verbal communication through headphones connected to the headsets.<br>- VE was a VR experience in an entertainment centre; participants could walk freely through a 3D modelized space station, interacting with the VE using their hands; only the duos were allowed to interact (verbal and physical touch) with each other. | Multimedia entertainment centre | Co-located | One session for 38 minutes for each participant | Participants were screened for motion sickness propensity, but all were retained. |
| Social-VR Program (Kalantari et al., 2023) | - Oculus Quest 2 headset with controllers; microphone and speaker for transmitting sound through Zoom.<br>- VEs for three modules were developed using the Spatial platform, and for one module pre-made 360-degree videos available in the Alcove VR app was used. | Lab sites (at three different locations) | Remote (each participant was paired with a partner from a different city) | One session for about 40-50 minutes for each pair of participants | Motion sickness was screened as an exclusion criterion and post-experiment questionnaires: There was some simulator sickness reported, but the rates were low (M = 21.82 out of 235.62; SD = 26.69). |
| Social VR-based Exergame (Played collaboratively [PC] group; Shah et al., 2023) | - Meta Quest headset.<br>- VE created using off-the-shelf assets of Unity3D and the Simple-Nature-Pack asset tool; Mixed Reality Toolkit (MRTK) was used to create the user interface and to integrate the interaction mechanisms with user interface controls; use of direct hand manipulation was made possible through the Oculus Integration Software Development Kit (SDK) for hand tracking along with MRTK; multiplayer options supporting a distributed connection over the wireless network between game players and networked information implemented using Normcore SDK. | Rehabilitation centre | Remote (different rooms of the same building) | Each game session for 10-15 minutes. Five-week study: each participant played the exergame twice a week for a total of 99 exergaming sessions for all participants. | Simulator sickness was evaluated before and after the user's exposure. The total score (TS) and aggregate of subfactors (O, N, D) were calculated for preexposure, postexposure and change in score: pretest (9.08 ± 11.19), post-test (14.43 ± 11.29), change (5.34 ± 6.74). Large difference on the item 'sweating' on the subscale nausea, indicating that the players were physically engaged in the exergame. |
| Motor Skill Learning (Virtual co-embodiment group; Kodama et al., 2023) | - Meta Quest 2 headset with controllers.<br>- VE was developed using Unity; positions and postures calculated for each frame were reflected in the co-embodied avatar using the Final IK Unity package. | University lab | Co-located | One session for a total experimental time of about 45 minutes for each participant | None reported |
| The Island (Experimental group; Barberia et al., 2018) | - Oculus DK2 headset with its internal head-tracking used for head rotation and orientation; application executed on 4 PCs (one was a server) each with a Kinect 2 sensor to capture upper body | Not specified | Remote (physically located in separate adjoining rooms) | Six sessions, one session per day for six consecutive working days for each participant. The sessions on | A question (asked during the screening session after a short VR experience) checked that participants did not feel dizzy while in the VE and the maximum |
| | movements of seated participants and gesture recognition.<br>- VE was created starting from the demo 'tropical paradise' released in 2007 with 'Unity3D 2.x' and then upgraded and heavily modified for rendering in VR; Virtual Bodies created using DAZ3D system and exported to Maya. Custom software included C# scripts to control the evolution of the body in order to produce smooth transitions from the child to the elderly version. | | | the virtual Island lasted for about 13.5 minute. | accepted level of dizziness was 3 (out of the maximum of 7). During the experiment one participant was invited to end her participation due to her discomfort due to simulator sickness during sessions. |
| MUVR Remote Psychotherapy (Matsangidou et al., 2022) | - Oculus Rift headset with its head-tracking to stream audio and visual content, and controllers; each headset paired to two Oculus Sensors that captured the user's physical position and movements.<br>- VE was developed using Unity3D and SteamVR; 3D models (human body and the VE) were created in Adobe Fuse CC and Maya, enhanced by Unity Assets; buttons on the controller were marked in the VE with different colours to improve the identification of each corresponding task; Salsa Lip-Sync RandomEyes and Photon Voice were integrated to allow the synchronous verbal communication; Audio dialogue files were processed in real time to automate the lip synchronization process and animate the avatar's facial expression; Photon Unity Networking was used to implement multiuser capability (up to 20 simultaneous users). | Laboratory | Remote (in two separate rooms within the same building) | One session for each participant (therapeutic session lasted approximately one hour and the entire trial lasted approximately two hours) | None reported |

## Applications

*Wellbeing Promotion.* While the reviewed studies had various research aims and objectives, the majority (seven out of ten) focused on well-being promotion as the primary health aim (see Table 3 and Figure 2). Among these, four studies utilised the ICVEs to facilitate social interactions between participants in order to enhance motivation and engagement in physical rehabilitation (Høeg et al., 2023; Shah et al., 2023), support group reminiscence (Baker et al., 2021), and produce meaningful social experience and improve mood (Kalantari et al., 2023), all among older adults. Furthermore, two studies leveraged SVEs to explore and positively impact affect among people in recovery from substance use disorders through mental health peer support events (Robinson et al., 2023), and another one in the context of a multimedia entertainment centre with real-world partners and friends as participants (Guertin-Lahoud et al., 2023). Moreover, Barberia and colleagues (2018) created a simulation of a complete life cycle (from birth to death and post-death out-of-body experience) via ageing avatars to explore its impacts on participants' death anxiety and changed attitudes to life.

*Symptom Reduction.* Besides well-being promotion, two studies incorporated established mental health therapeutic approaches into their ICVEs to achieve some degrees of symptom reduction among their participants. Namely, Li and Yip (2023) integrated an arts therapy approach (through 3D drawing) with a therapist in their CVE to reduce perceived stress and improve mental well-being among young adults with moderate to high stress levels. Another study applied principles and techniques of three psychotherapeutic approaches (such as exposure therapy) to their VEs with female students at high-risk for an eating disorder (Matsangidou et al., 2022).

*Skill Acquisition.* The remaining research focused on skill acquisition as its main health aim. Kodama and colleagues (2023) explored a virtual co-embodiment training approach, with both trainer and participant being embodied in one avatar, to enhance motor skill learning efficiency in a dual task.

*Table 3. Quality assessment and outcome scales of the included studies; *JBI Critical Appraisal Checklist for Quasi-Experimental Studies; ^JBI Critical Appraisal Checklist for Qualitative Research; ~ For the full reference of citations inside the table refer to the corresponding study.*

| Name of Program (Name of ICVE Intervention Group, if multiple groups) | Quality Assessment | | Type of Comparator Group/s (if any) | Health Aim/s | Health Outcome Variable Scale/s | Feasibility Outcome Variable Scale/s (including presence and copresence) |
|---|---|---|---|---|---|---|
| | Quasi-Experim.* | Qual.^ | | | | |
| School Days (Baker et al., 2021) | N/A | 1.0 | N/A | **Well-being promotion** (to have a positive impact on the lives of older adults) | Semi-structured interview | -Adapted version of Witmer and Singer's presence questionnaire (1998)~<br>-Semi-structured interview<br>-Observation notes, photographs and short video recordings |
| Buddy Biking (Høeg et al., 2023) | 0.60 | 0.40 | N/A | **Well-being promotion** (to facilitate social interaction for VR-based rehabilitation purposes) | Interpersonal interaction questionnaire for observers (IPIQ-O; Goršič et al., 2019b) | -Simulator sickness questionnaire (SSQ; Kennedy et al., 1993)<br>-System usability scale (SUS; Brooke et al., 1996)<br>-Intrinsic motivation inventory (IMI; Ryan & Deci, 2000)<br>-Virtual embodiment questionnaire (VEQ; Roth & Latoschik, 2020) |
| Cognitive Behavioral Immersion (Robinson et al., 2023) | 0.67 | 0.70 | N/A | **Well-being promotion** (change in affect for people in recovery from SUDs) and **Skill acquisition** (to facilitate cognitive-behavioral skill acquisition) | Positive and negative affect schedule (PANAS; Watson et al., 1988) | -Net promoter score (NPS; Reichheld, 2003)<br>-Online social support scale (OSSS; Nick et al., 2018)<br>-Structured interviews |
| Remote Arts Therapy (Li & Yip, 2023) | N/A | 0.70 | N/A | **Well-being promotion** (enhancing mental well-being) and | -Warwick-Edinburgh Mental Wellbeing Scale | -System Usability Scale (SUS; Brooke, 1996)<br>-Copresence (Nowak & Biocca, 2003) |
| | | | | **Symptom reduction** (reducing perceived stress) | (WEMWBS; Tennant et al., 2007)<br>-Perceived Stress Scale (PSS; Cohen et al., 1983)<br>-Semi-structured interview | -Semi-structured interview |
| International Space Station (Duo group; Guertin-Lahoud et al., 2023) | 0.78 | 0.50 | Immersed in VE without interaction with others (Solo group) | **Well-being promotion** (positive affect) and **Symptom reduction** (state anxiety) | -Positive affect (Game experience questionnaire, Positive affect component; IJsselsteijn et al., 2013)<br>-State anxiety (STAIS-5) | -Copresence (Poeschl and Doering's Co-presence subscale; Witmer & Singer, 1998)<br>-Presence (Slater–Usoh–Steed presence questionnaire; Usoh et al., 2000)<br>-Immersion (Presence questionnaire, Adaptation/immersion subscale; Witmer et al., 2005) and sensory immersion (Game experience questionnaire, Sensory/imaginative immersion component; IJsselsteijn et al., 2013)<br>-Flow (Flow component; IJsselsteijn et al., 2013)<br>-Psychophysiological and motion sensors<br>-Semi-structured interview |
| Social-VR Program (Kalantari et al., 2023) | 0.89 | 0.80 | N/A | **Well-being promotion** (meaningful social experience, enhancing mood) | -Perceived social presence (Nowak & Biocca, 2003) and Likeliness to Reconnect with Partner (Boothby et al., 2018), as well as post-immersion interviews<br>-Multidimensional Mood State Questionnaire (Steyer et al., 1997) | -Self-Assessment Manikin (SAM; Bradley & Lang, 1994)<br>-Acceptance of Head-mounted Virtual Reality in Older Adults (Huygelier et al., 2019)<br>-Motion sickness (Kennedy et al., 1993), excessive cognitive workload (Hart & Staveland, 1988) and Usability Metric (Finstad, 2010)<br>-Qualitative Interview |
| Social VR-based Exergame (Played collaboratively [PC] group; Shah et al., 2023) | 1.0 | 0.90 | Immersed in VE individually (Played alone [PA] group) | **Well-being promotion** (to facilitate full-body exercise and social collaboration) | -Intrinsic motivation inventory (IMI; Ryan & Deci, 2000) | -System Usability Scale (SUS; Brooke, 1996)<br>-Simulator sickness questionnaire (SSQ; Kennedy et al., 1993) |

Systematic Review on Immersive Collaborative Virtual Environments in Health

| | | | | | -Physical exertion (total distance travelled by the user's hands)<br>-Qualitative feedback (collaborative meetings, field observations, interviews and a questionnaire based on text input) | -Game experience questionnaire (GEQ; Poels et al., 2007)<br>-Virtual embodiment questionnaire (VEQ; Roth & Latoschik, 2020) |
|---|---|---|---|---|---|---|
| Motor Skill Learning (Virtual co-embodiment group; Kodama et al., 2023) | 1.0 | N/A | Two groups: Immersed in VE individually (Alone group) and Immersed while sharing 1P perspective with trainer (Perspective-sharing group) | **Skill acquisition** (motor learning efficiency) | Task performance (Number of correct hits) | Virtual embodiment questionnaire (VEQ; Roth & Latoschik, 2020) |
| The Island (Experimental group; Barberia et al., 2018) | 1.0 | N/A | Waiting list (Control group) | **Well-being promotion** (death anxiety and changed attitudes to life) | -Collett-Lester Fear of Death Scale (Lester & Abdel-Khalek, 2010)<br>-Life-Changes Inventory (Greyson & Ring, 2004) | Customized post-trial questionnaire |
| MUVR Remote Psychotherapy (Matsangidou et al., 2022) | N/A | 0.90 | N/A | **Symptom reduction** (Acceptance and Commitment Therapy, Play Therapy and Exposure Therapy for sufferers with body shape and weight concerns) | Semi-structured interviews | -System Usability Scale (SUS; Brooke, 1996)<br>-Presence (Nichols et al., 2000)<br>-Qualitative observational notes and video/audio recordings<br>-Semi-structured interviews |

## Health & Feasibility Outcomes

### Facilitating engagement in physical rehabilitation

Both studies that aimed at facilitating social interactions by using ICVEs to improve older adults' engagement in physical rehabilitation reported positive results.

*Buddy Biking*. Høeg and colleagues (2023) measured the amount and quality of interactions between participants cycling on a virtual tandem bike using an observer rating scale. They reported a positive game-related conversation between co-players, with those older adults who biked with a family-member or friend rather than a researcher showing higher amount and balance of conversation. With respect to feasibility scales, they found high average motivation scores in all categories of interest/enjoyment (6.5 ± .7), effort/importance (6.1 ± .7) and relatedness (6.4 ± .6), with the score of seven corresponding to 'very true' on relevant questions. Also, 75% of participants somewhat- to strongly agreed on items inquiring about their perceived ownership of the virtual body (e.g., 'felt like the virtual body was my body'). The program's average usability score was excellent (85 ± 5 out of 100; Bangor et al., 2009). In post-experience interviews, the majority of participants described it as enjoyable and engaging, with all except one (who experienced dizziness and had to end the session prematurely) expressing an interest to use the system once more as part of their therapy. Additionally, they corroborated that teamwork and collaboration were encouraged by the Buddy Biking experience and that it enabled distraction and reduced the perceived cycling time. Participants also reported some areas for improvement, such as the mismatch between physical input and virtual feedback.

*Social VR-based Exergame*. Shah and colleagues (2023) compared motivation and physical exertion in both within- and between-subjects designs among older adults in a virtual fruit-picking task. Comparison in motivation scores between solo versus collaborative conditions showed significantly higher scores in all motivation categories for those in the collaborative group (interest/enjoyment: *t*[12] = 2.94, *p* = .02; effort/importance: *t*[12] = 4.01, *p* = .002; value/usefulness: *t*[12] = 2.82, *p* = .01). Within-group comparison, enabled by switching between conditions for the last session, demonstrated significant improvement in the motivation scores of participants in the solo condition switching to the collaborative condition, with medium to large effect sizes in different categories (Cohen's *d* ranging from .69 in interest/enjoyment to .98 in effort/importance). Conversely, significant deterioration was reported among those switching from the collaborative to the solo condition (Cohen's *d* ranging from .71 in value/usefulness to .91 in interest/enjoyment). Between-group comparison of physical exertion showed the similar trend of significantly higher scores in the collaborative condition (*t*[12] = 7.09, *p* < .001). Similar to motivation, switching from solo to collaborative condition increased exertion scores (Cohen's *d* = .98), and converse switching reduced them (Cohen's *d* = − .72). Considering the feasibility measures, the usability of the program was scored as excellent, with an average of 83.75 ± 13.3. Furthermore, on the perception of ownership of the virtual body, 71.4% strongly agreed on relevant items. Reflecting on their experiences through qualitative feedback, participants described the intervention as engaging and enjoyable. They found the social and collaborative aspects to enhance the physical health benefits. Additionally, they emphasised the motivational effect of the desire to perform better in each session, noting that participants in the collaborative condition tended to maintain motivation longer than those in the solo condition, even when facing losses during the exergaming sessions.

Finally, participants perceived the program as offering various cognitive and physical health benefits, including improved mobility, activity level, balance, eye-hand coordination, and attention span.

**Supporting meaningful social interactions**

Using ICVEs to support meaningful social interactions among older adults who were physically remote from each other, proved to be fruitful in two studies.

*School Days*. As part of a qualitative research, Baker and colleagues (2021) created a social VR experience for older adults to support group reminiscence among them and positively impact their lives. This was enabled in the virtual context of a modern classroom and through the use of three sets of integrated mechanics as social lubricants, namely, conversation starters, personal artifacts, and Avacasts. In post-experience interviews, the majority of participants (77% of the 66 responses) were in agreement that the VE (i.e., the virtual classroom) was engaging for them in terms of visual information. Additionally, it was effective in surfacing memories and facilitating reminiscence. Conversation starters and personal artifacts were useful in generating school stories and scaffolding closer social bonds between participants. This was exemplified, through video footage analysis, in how a single conversation starter stimulated reminiscing about different aspects of participants' lives for over 27 minutes in one session. Avacasts, as holograms telling the stories of non-present people using their pre-recorded audios, were an attempt to introduce a new reminiscence-supporting feature without increasing the complexity of the environment by adding further real-time users. It helped inspire participants to respond to these stories and reflect on their own experiences. In terms of usability considerations, some older adults struggled to grasp the required input of the controllers for the corresponding hand movement gestures of the avatars. Moreover, video data of some participants revealed that they desired the VE and its elements to be more dynamic, for example, the clock showing the real-world time rather than being fixed.

*Social-VR Program*. Kalantari and colleagues (2023) aimed to facilitate meaningful social interactions among pairs of older adults from different cities using ICVEs and 360-degree videos of tourist destinations. Post-experiment measures indicated moderately high ratings of Perceived Social Presence in the VE ($M$ = 61.21 ± 22 out of 100) and Likeness to Reconnect with the VR partner ($M$ = 3.69 ± .79 out of 5). The only significant mood state change observed from pre- to post-program was on the calm-nervous dimension, with an average 3-point shift towards the 'calm' state ($SD$ = 6.93; $t[17]$ = 2.53, $p$ = .022). Feasibility measures revealed a high level of Engagement in the VE ($M$ = 4.18 ± .91 out of 5), moderately high degrees of spatial presence on subscales of Possible Actions ($M$ = 3.70 ± 1.17 out of 5) and Self-Location ($M$ = 3.76 ± 1.27 out of 5), and a high average usability score among participants ($M$ = 67.01 ± 20.73 out of 100). Analysis of interview findings showed predominantly positive affective responses (81 positive vs. 10 negative statements), with participants describing their experiences as 'comfortable' and 'interesting'. Positive mood changes, attributed to an increasing sense of mastery in VR, were reported in all relevant responses (8 bad to good mood vs. 0 reverse direction). However, feelings of social presence elicited mixed responses (25 present vs. 24 non-present), with challenges such as the inability to gauge partner's emotions via gestures being highlighted. Notably, a pairing effect was observed in most pairs, with partners expressing similarly valanced responses. Some participants identified obstacles to spatial presence (28 present vs. 12 non-present), including the inability to move voluntarily in the 360-degree video section and the inability to see their own avatars.

Recommendations for improvement included incorporating a partner-matching component, allocating more training time, using wireless and more user-friendly headsets, and simplifying controller design.

**Promoting positive affect**

In two different research designs across two studies — one comparing pre- vs. post-experience and the other comparing solo vs. duo conditions — positive affect was found to be significantly heightened in the post-experience and duo condition, respectively.

*Cognitive Behavioral Immersion (CBI)*. CBI is a program designed to deliver cognitive-behavioural therapy (CBT) techniques by lay coaches in peer support events via a metaverse application called *Innerworld* (Robinson et al., 2023). In this study, the authors evaluated their program's potential for individual users self-identified as in recovery from a substance use disorder. A significant increase in positive affect was reported ($t[19] = 2.76$, $p = .01$, Hedges' $g = .31$) among 20 participants during their most recent CBI session from pre- ($M = 3.03$, $SD = .7$) to post-session ($M = 3.32$, $SD = .75$). Among the feasibility scales, user engagement with the application was found to be high; participants frequented the platform for an average of 19.34 hours ($SD = 43.96$; range = .4–291), with Net Promoter satisfaction score of 66.67 being in the excellent range. In structured interviews with 11 participants, eight themes were identified and reported. Regarding technological usability, the most common responses were on difficulty navigating VR for the first time (18 comments), and some aspects of the user experience interface lacking the needed visual social cues for users (11 comments). In terms of the sense of community, 30 comments centred around shared experiences and emotional connection with others in the program, while 23 responses valued the online community due to the diversity of members. Furthermore, in comparing CBI with other mental health interventions, participants were in agreement that it could be helpful as an additional resource to people's recovery journey. With respect to the psychological impacts of the program, participants most frequently highlighted improvements in their wellbeing (33 comments) and positive emotions (18 comments). The other four themes were less frequently mentioned by users, but expressed challenges with recovery sessions, the influence of anonymity in sessions, feelings of presence and immersion, and COVID-19 pandemic impacts.

*International Space Station.* Utilising the real-world context of a multimedia entertainment centre, Guertin-Lahoud and colleagues (2023) explored differences in the perceived and lived experience of two groups of participants. One group engaged with the International Space Station VR experience individually, while the other interacted in dyads with real-life partners and friends. They found significantly higher positive affect scores in the duo group ($M = 6.489$, $SD = .718$) compared to the solo group ($M = 5.608$, $SD = 1.086$; $z = -2.491$, $p = .01$). Regarding feasibility measures, the duo group reported significantly higher levels of copresence ($z = 2.722$, $p = .005$), while no significant difference was observed in presence and immersive experiences. Interestingly, the interactivity of the VR experience (i.e., active exploration vs. passive 360-degree video watching) impacted the sense of adaptive immersion differently in the two conditions. Solo participants experienced heightened adaptive immersion during the passive experience, while duos experienced it during the active VR exploration. In post-experience semi-structured interviews, 20 out of 28 participants expressed a preference for engaging in the experience in dyads, or indicated that they would have preferred it if they were in the solo group. Additionally, the majority of participants (16 out of 28) reported a higher sense of presence during the active phase of the experience. Lastly, 12 participants noted feeling more excited and awake post-program compared to their pre-program state, although some reported experiencing physical strain as well.

**Simulating unique life experiences**

*The Island*. Exploring more unique experiences (i.e., virtual mortality and near-death experience) using SVEs, proved to result in positive changes in life-attitude among female young adults. Barberia and colleagues (2018) created an otherworldly SVE in the virtual setting of an *island* to investigate the effects of simulated mortality/immortality on various outcomes, including death anxiety and attitudes toward life. Utilising a Bayesian model, they demonstrated the significant effect of condition (i.e., experimental group vs. waiting-list control group) on changed attitudes toward life ($P[\beta>,0] = .992$), but not on death anxiety ($P[\beta>,0] = .632$). Exploring further, they reported significantly higher scores on Life Changes inventory for those in the experimental group ($M = .47 \pm .071$), compared to the control group ($M = .25 \pm .046$; Cohen's $d = -.91$). With respect to the feasibility measures, benefitting from a customised presence and copresence scale, they found strong senses of being there and being with others in the VE, and perceiving the events as really happening among participants. Additionally, participants experienced high senses of body ownership and agency over their avatars, despite the 'ageing' nature of their virtual bodies.

**Leveraging mental health therapies**

In the two studies that integrated mental health therapeutic approaches in their ICVEs, qualitative findings predominantly suggested positive changes in participants, indicating the potential for further research in this emerging field.

*Remote Arts Therapy.* Li and Yip (2023) utilised an ICVE in a pilot case study to assess the potential synergy between arts therapy and VR technology through eight therapy sessions with three young adults with moderate to high levels of stress. The findings of two quantitative scales measuring changes of wellbeing and stress from pre- to post-program showed no consistent trend, with only small changes in scores. All the while, in semi-structured interviews, participants reported changes in their feelings and emotions following the program. These changes included expelling negative emotions through sessions, having a chance to reflect on thoughts and emotions, providing a sense of fulfillment and positivity, and helping them to get calm and focused. Regarding the feasibility measures, usability scores ranged from good to excellent (75-90 out of 100). Additionally, session notes from the therapist revealed several noteworthy experiences reported by participants, including an increasing bond with the therapist over the course of sessions, a strong sense of freedom while drawing in 3D space, heightened inward and outward exploration following initial sessions, a feeling of absorption in the creation of artwork, and a sense of 'thrill' upon completing it.

*MUVE Remote Psychotherapy*. As part of a codesigned and multidisciplinary project, Matsangidou and colleagues (2022) developed MUVEs for remote psychotherapy. The qualitative research examined the acceptability of these VEs among female participants with body weight and shape concerns, and therapists. Following the analysis of semi-structured interviews with participants and detailed observation notes made by the researchers throughout the sessions, four core themes emerged. In the theme of Remote Psychotherapy, all seven therapists and 11 out of 14 participants reported experiencing a sense of trust and security. This was reflected through the disclosure of thoughts, emotions, and personal information by participants. The lack of face-to-face communication, as expressed by all therapists and eight participants, along with the cartoony cube-like avatar of the therapists, were identified as contributing factors to the sense of safety and self-acceptance. In relation to the two additional themes, various patterns emerged, including the necessity for a diverse range of activities tailored to participants' interests, the creation of a relaxed and informal atmosphere through the use of games and gamification elements, ensuring the therapy experience is both believable and realistic, and the content and process of therapy eliciting emotions while enabling therapists to identify and respond appropriately to them. In terms of feasibility measures, excellent average rates of system usability (81.5%) and presence (5.15 ± .95 out of 7) were reported by both participants and therapists. Furthermore, the last theme of the qualitative analysis, which focused on technological aspects, supported these scores and suggested that users were open to the idea of using MUVE therapy in the future. Nonetheless, both participants and therapists, as well as the researchers observing them, emphasised the need for familiarisation with VR technology and the user interface before initiating therapy. Ultimately, participants noted that virtual therapy was beneficial in helping them accept and foster more positive attitudes toward their bodies.

**Optimising motor skill learning**

*Virtual Co-embodiment*. Using a novel method (i.e., virtual co-embodiment) in motor skill learning within an ICVE, Kodama and colleagues (2023) demonstrated its higher efficiency compared to two other conditions in a randomised controlled trial. After establishing that skill learning improved throughout the trials across all three conditions (Virtual Co-embodiment, Perspective Sharing, and Alone), post-hoc analysis revealed a significantly greater improvement in the Co-embodiment group compared to the other conditions ($p < .001$ in all learning trials). Notably, while participants in the Co-embodiment group also experienced the highest performance drop during the test phase ($p < .0001$, between all possible pairs), they still scored significantly higher than those in the Perspective Sharing ($t[60] = 3.18$, $p < .05$) and Alone conditions ($t[60] = 2.23$, $p < .05$). In relation to feasibility measures, no significant difference was reported between conditions regarding the Sense of Embodiment, including Agency and Body Ownership.

## Design Characteristics of the ICVEs

Table 4 and Figure 2 provide an overview of design features embedded in the ICVEs. Various virtual settings were employed across studies, with four of them mainly inspired by natural landscapes (Barberia et al., 2018; Høeg et al., 2023; Matsangidou et al., 2022; Shah et al., 2023). The number of simultaneous users in the same VE was more consistent, with the majority of studies (seven out of 10) allowing for two active participants (Guertin-Lahoud et al., 2023; Høeg et al., 2023; Kalantari et al., 2023; Kodama et al., 2023; Li & Yip, 2023; Matsangidou et al., 2022; Shah et al., 2023). Notably, one study did not have any limitations in terms of the number of simultaneous users, as it was conducted over an MMO metaverse application (Robinson et al., 2023). The most common modes of embodied interaction among participants via avatars were verbal communication, gestures and shared activities. Additionally, two studies incorporated facial expressions (Baker et al., 2021; Matsangidou et al., 2022). Across all studies, the presence of facilitators or moderators alongside participants or within VEs provided a consistent support before, during, and/or after VR sessions.

Complementing the identified design characteristics, ten design factors worth considering in evaluating the affordances of various CVEs, as elaborated in Dalgarno and Lee (2010), are reported in Table 5 for each study. Notably, all ICVEs provided varying degrees of the smooth display of view changes and object motion, consistency of object behaviour, and embodied actions, while none afforded the control of environment attributes and behaviour to participants.

*Table 4. Virtual contexts and design features of the ICVE interventions.*

| Name of ICVE Intervention | Virtual Setting | Number of Simultaneously Immersed Users | Modes of Interaction | Task or Objective | Provided Support |
|---|---|---|---|---|---|
| School Days (Baker et al., 2021) | The Hall (to explore the system's functionality and choosing an avatar while looking at a virtual mirror) and The Classroom (a room styled after a modern school environment with desks, chairs and various paraphernalia [stationery, a world map, a globe]) | Groups of 2–3 (plus a facilitator) | Verbal interaction, facial and lip movements, gestures and shared endeavour via avatars | The participants used School Days to visit a virtual classroom where they could reminisce about their school experiences via three main design mechanics: conversation starters, personal artefacts, and Avacasts (hologram storytellers) | -Designed to be used in a seated position to reduce simulator sickness and to support participants with mobility difficulties.<br>-Designed to incorporate the role of a facilitator inside the virtual environment (to guide them through social experiences, and to provide in situ technical support).<br>-Members of the research team were always present alongside each participant.<br>-The Hall provides an opportunity for users to become familiar with the virtual environment, introduce themselves to one another, and explore the system's functionality, such as by passing virtual objects to each other.<br>-The desk [in The Classroom] has another sphere that allows users to teleport back to The Hall if desired. |
| Buddy Biking (Høeg et al., 2023) | A shared virtual tandem bike, situated in a high-altitude mountainous environment on a | Two | Shared endeavour, and verbal interaction not embodied in VE | A two-user biking challenge that placed users together on a shared virtual tandem bike to travel the virtual gravel path, which formed a looping circuit. | A short introduction of potential side effects they might experience during VR-exposure. |
| | gravel path with flora and fauna | | | Users shared the experiences of anticipating ascents and conquering them together. | |
| Cognitive Behavioral Immersion (Robinson et al.,2023) | An immersive MMO app called Innerworld comprising a collection of different worlds (ranging from a home base to environments that depicted other types of scenes, e.g., office, campground, outer space) | Not limited (varied) | Verbal interaction, gestures, and shared endeavour via avatars | The core game loop is social interaction with an opportunity to choose from a list of upcoming mental health events (e.g., check-in groups and topic-focused meetings) and recreational events (e.g., drawing games, chess tournaments, charades). | -Innerworld provides an introductory tutorial to teach individuals how to navigate the application<br>-Trained lay coaches facilitated events |
| Remote Arts Therapy (Li & Yip, 2023) | Replica of a minimal room | Two (including therapist) | Verbal interaction, gestures and collaborative art creation (drawing) via avatars | CVE-enabled remote arts therapy: each session's art creation theme and activities were designed to be different, but they were all aligned with psychotherapeutic objectives following pre-defined protocols | The experimenter demonstrated how to use the enabling software and hardware. |
| International Space Station (Guertin-Lahoud et al., 2023) | 3D modelized International Space Station (ISS) representation | Two (among strangers without interaction with them) | Verbal interaction (only between users in dyads) and gestures via avatars | ISS exploration, watching 360 videos of astronauts and watching unnarrated rotation around planet earth while seated | -Potential technical issues (e.g., low battery and erroneous tracking of headsets) and participants' progression through the VR experience were monitored in real time on the moderator's tablet.<br>-Instructions on virtual space navigation were provided |
| Social-VR Program (Kalantari et al., 2023) | Small room with window view (Training Module), a room with a large world map in the | Two active participants (plus a moderator) | Verbal interaction, gestures and shared endeavour via avatars | Practicing and navigating VR, visiting a tourist destination and creating a photo collage design together to facilitate social connection. | -During sessions two researchers were present at each site: one for administering the questionnaires and monitoring safety issues, and one for coordinating the |
| | middle (Introductions Module), 360-degree videos of tourist destinations (Travel Module), a gallery space (Productive Engagement Module) | | | | technological setup and served as a moderator within the virtual environment.<br>-Participants sat in a swivel chair or stood as desired throughout activities.<br>-A 5-min break was given after each module. |
| Social VR-based Exergame (Shah et al., 2023) | Nature-based environment | Two | Verbal interaction, gestures and shared endeavour through avatars | A fruit-picking game and social collaboration through team-based game tasks and reward mechanisms | -A demo of the exergame was provided and participants were asked to inform physiotherapists about any discomfort during the sessions.<br>-A familiarization session was then held in which participants experienced the flow of the exergame to reduce potential learning effects. |
| Virtual Co-embodiment (Kodama et al., 2023) | A generic room | Two (co-embodied in one avatar) | Shared endeavour (same weighted control of avatar's hand movements) | The task included the simultaneous drawing of a seven-pointed star with the right hand and a five-pointed star with the left hand. | -A tutorial for the dual task and explanation of system.<br>-Removing the headsets and resting for 3 minutes. |
| The Island (Barberia et al., 2018) | A room with large mirrors (Tutorial), and main experiment in an alien island covered with colourful vegetation, mountains and bridges, with day-night cycles | Three | Gestures and shared endeavour via avatars | Providing a first-person experience of a life cycle that simulates aspects of birth, childhood, maturity, decay, death, transition and post-death survival (through a simulated near-death and out-of-body experience) via ageing avatars. Reinforcing the social bond between participants through | -Presence of two confederates in the role of 'older' participants in the beginning.<br>-An experimenter assigned to each participant.<br>-A 2-minute voice tutorial for teaching navigation in the virtual world.<br>-A second tutorial to teach how to interact with objects. |
| | demonstrated by sunrises and sunsets. | | | three collaborative tasks that required cooperation between them (classification task, maze task, piano task). | |
| MUVR Remote Psychotherapy (Matsangidou et al., 2022) | Natural landscapes (e.g., desert, forest) and a room with a mirror | Two (including therapist) | Verbal interaction, facial expression, gestures and shared endeavour via avatars | Full session consisted of three stages: 1) Tutorial; 2) Starting location where the therapist greets the participant and participants create their virtual avatar; 3) Three VEs (2 Acceptance Commitment Therapy values VEs or 2 games, and a Mirror Exposure VE). | -A tutorial of about 25 minutes in order to become familiar with the use of the VR system<br>-Two researchers were present with participant and therapist (in separate rooms) |

## Quality Ratings

Following the assessment of included studies by two independent reviewers, quality ratings ranged from 0.40 to a maximum of 1.0 (see Table 3). The seven studies evaluated against items on the Quasi-Experimental Studies checklist (average score = .85) tended to stop short of receiving higher scores based on the lack of a control group and multiple measurements of outcomes both pre- and post-intervention. On the other hand, the unclear state of the philosophical perspective (and its congruity with the research methodology), the absence of a statement locating the researcher culturally or theoretically, and the unaddressed influence of the researcher on the research were amongst the most common factors keeping the eight assessed studies from scoring higher on the Qualitative Research checklist (average score = .74).

*Table 5. Evaluation of the reviewed studies and their ICVEs against identified design factors, based on Dalgarno and Lee (2010).*

| Name of ICVE Intervention | Realistic display of environment | Smooth display of view changes and object motion | Consistency of object behaviour | User representation | Spatial audio | Kinaesthetic and tactile force feedback | Embodied actions | Embodied verbal and non-verbal communication | Control of environment attributes and behaviour | Construction/scripting of objects and behaviours |
|---|---|---|---|---|---|---|---|---|---|---|
| School Days | Yes | Yes | Yes | Yes | Yes | No | Yes | Yes | No | No |
| Buddy Biking | Yes | Yes | Yes | Partly | Partly | Yes | Partly | No | No | No |
| Cognitive Behavioral Immersion | Partly | Partly | Yes | Yes | Yes | No | Yes | Yes | No | No |
| Remote Arts Therapy | Yes | Yes | Yes | Partly | Yes | No | Yes | Yes | No | Partly |
| International Space Station | Yes | Yes | Yes | Partly | Yes | No | Yes | Yes | No | No |
| Social-VR Program | Yes | Yes | Yes | Yes | Partly | No | Yes | Yes | No | No |
| Social VR-based Exergame | Partly | Yes | Yes | Partly | Yes | No | Yes | Yes | No | No |
| Virtual Co-embodiment | Partly | Yes | Yes | Partly | No | Yes | Partly | Partly | No | No |
| The Island | Partly | Yes | Yes | Yes | No | No | Yes | Partly | No | No |
| MUVR Remote Psychotherapy | Yes | Yes | Yes | Yes | Yes | No | Yes | Yes | No | No |

## Discussion & Implications

In the literature on the applications of immersive technologies, there is an observation of a push-pull phenomenon that has been suggested to accelerate the adoption of these technologies into diverse applications, including in the health realm (e.g., Steed & Schroeder, 2015; Weiss & Klinger, 2009). The 'push' arises from continuous advancement in XR hardware and software technologies, while the 'pull' stems from the needs of clients and service providers for more efficacious, accessible, and tailored interventions. The strain of the COVID-19 pandemic on the health sector and the ensuing increasing demand for non-face-to-face services have further fuelled both sides of this phenomenon (Lee, 2022).

This systematic review aimed to synthesise and report on the current state of the literature regarding the emerging use of immersive shared environments in health applications. Ten studies were identified and included in this review, conducted in various countries and with different populations, scoring high average research quality ratings for qualitative (74%) and quantitative (85%) data analyses. Among these, seven were experimental research, including two RCTs. In the following sections, discussion points pertaining to different aspects of the included studies are unfolded, elaborated upon, and compared.

### Populations

Included studies were almost entirely conducted with two age populations: young adults and older adults. As older adults tend to be overlooked in interventions based on technological innovations, the use of shared XR technologies is promising but also comes with certain caveats for this population. On one hand, it enables older adults to connect with relatives, friends, and health professionals, keeping them motivated to be more active. This benefit was directly demonstrated in two studies in this review in the case of physical rehabilitation (Høeg et al., 2023; Shah et al., 2023), while another two studies explored the possibility of making meaningful social connections for older adults despite physical distance by using social scaffolds (Baker et al., 2021; Kalantari et al., 2023).

However, based on the reviewed articles, special attention needs to be paid to making the technological setup and user interface as compatible as possible with older adults' needs and capabilities, as they tend to be generally frailer and less technology proficient. This can be achieved by headset manufacturing companies creating less complex and more comfortable devices for older adults (e.g., lighter headsets with simpler controllers), and by health programs' creators planning for the co-production of their ICVEs with participants and allocating specific training time with the technology and the interface as necessary steps towards achieving their health outcomes.

Also relevant to the discussion of target populations in XR health studies is the need to account for individual differences among participants on various levels, such as secondary health conditions, personality traits, and attitudes towards technology. This was reflected in the possible influence of different learning styles of participants in a skill acquisition study (Kodama et al., 2023), and in underlying mild cognitive impairment among some older adults and its relationship with the sense of presence in the VE (Kalantari et al., 2023). These considerations have important implications for determining the optimal degree of immersion, design of the ICVE and study tasks; as, for example, in the latter study authors suggested that inducing greater sense of spatial presence in the group of older adults with mild cognitive impairment could potentially increase the risk of them becoming 'lost' in the simulation.

## Interventions & Contexts

While our effort to capture studies using XR technologies other than VR (i.e., AR and MR) is reflected in the search string we developed (see review's protocol), no such study met this review's criteria. One MR research (Crowell et al., 2019) initially passed full-text screening but was later discovered during the data extraction stage to fall short of meeting the immersiveness criterion. This is noteworthy, as one distinctive promise of using multi-user AR and MR technologies for health applications is their potential to integrate sensory modalities beyond visual and auditory (such as touch and smell) into the participants' immersive experience (Bansal et al., 2022)—an advantage not yet fully realised in VR systems. As mentioned in the introduction, the number of modalities integrated into the system is one of the main factors affecting presence and copresence. Along this line, one included study (Høeg et al., 2023) used sensors on bicycle pedals to transform physical cadence into forward movement in the VE. In this setup, the steepness of the road in the VE required increasing cadence to maintain velocity.

Another aspect related to the technology of multi-user VR health interventions is the choice between using an existing platform, such as social VR applications, or creating a bespoke environment to host the health program. While the majority of the included studies created their tailored ICVEs, two studies were conducted on an MMO metaverse application (Innerworld; Robinson et al., 2023) and a social VR platform (Spatial; Kalantari et al., 2023). This decision depends on various factors such as study's aims, required affordances, budget, privacy issues, and copyright considerations, among others. However, considering the use of social VR platforms to create ICVEs that do not intend to gather sensitive personal information might lead to two sets of opportunities: first, for the research, by leveraging more sophisticated social scaffolds and mechanics already available on such platforms (McVeigh-Schultz et al., 2019); and second, for the dissemination and accessibility of health programs to a larger audience beyond the limited number of participants in a single study (for a comparison of different social VR platforms see Liu and Steed, 2021).

## Applications, Outcomes & Design Considerations

One noteworthy point observed across studies was the crucial role of qualitative findings in contextualising the interventions in terms of design, efficacy and feasibility, as well as in reporting insightful users' feedback regarding the technology and the intervention. One study (Baker et al., 2021) received the maximum score in the Qualitative Research quality assessment, reflecting their adoption of a stepwise comprehensive approach in methodology. At this stage of research in the field, there is still a need for user experience evaluations and qualitative analyses to unravel the network of technology-participant-health interrelationships (see Figure 2) before reaching an adequate standard for designing effective and engaging interventions based on known factors and their connections. As an example, in Baker and colleagues' (2021) study, older adults (participants element) expressed in the co-production process that they would prefer to reminisce in smaller groups (technology/design element) in order to make deeper meaningful connections (health element). This in turn prompted researchers to introduce holograms of non-present people using their pre-recorded audios as a creative reminiscence-supporting feature (technology/design element) without increasing the complexity of the environment by adding further real-time users.

Three included studies reported on direct comparisons between single- and multi-user engagement in their VEs regarding different outcomes. They demonstrated higher scores for the multi-user group in terms of positive affect and copresence (Guertin-Lahoud et al., 2023), motivation and physical exertion (Shah et al., 2023), and motor skill learning efficiency (Kodama et al., 2023) compared to the single-user group. Although each of these studies adopted distinct approaches in design, aim, and outcomes of interest, they collectively provide promising preliminary findings regarding the applications of ICVEs in health.

A caveat outlined in two studies (Høeg et al., 2023; Shah et al., 2023) and observed in several others was the potential confounding effect of the novelty of immersive technologies on study outcomes. For most of the population, we have yet to reach a point where the experience of 'being there together' in a VE feels as natural as talking on the phone. This concept, known as 'connected presence', varies across different new media, including SVEs, and relates to both presence and copresence. Connected presence is defined as the extent to which a medium or shared space, enabled by new technology, actually mediates everyday relationships by complementing face-to-face (i.e., unmediated) interactions (Schroeder, 2006).

To mitigate the novelty effect of experiencing ICVEs, two strategies were implemented in the included studies. One strategy, used by Barberia and colleagues (2018), involved multiple pre-program training sessions to help participants acclimatise to the 'new skin' of their avatars and gain a sense of mastery over the technology. Another strategy, applied by Shah and colleagues (2023), was to plan and conduct the program over multiple repeated sessions while assessing outcomes throughout. By considering and reducing the potential effect of users' excitement and/or anxiety about experiencing a novel technology on outcomes, the validity of health and feasibility findings would improve.

In the case of utilising ICVEs for mental health therapeutic sessions, two included studies (Li & Yip, 2023; Matsangidou et al., 2022) provided valuable implications for further research and practice. First, compared to previous experiences with videoconferencing platforms, the therapist in Li and Yip (2023) suggested that avatar-mediated communication via the ICVE was more relaxing and effective in establishing trust for both the participant and the therapist. Second, ICVEs enabled opportunities beyond the reach of face-to-face psychotherapeutic sessions, such as the client and therapist working on the same artwork in a 3D space simultaneously in the case of art therapy. Additionally, they offered the advantage of remaining anonymous and avoiding the fear of social stigma while communicating inner thoughts and feelings, without losing the chance to receive help from a real therapist. Despite the promising findings from these qualitative studies, future research needs to validate them with controlled trials and greater number of participants.

Most of the included studies attempted to facilitate collaboration and social connection between users in various ways. In the two studies that specifically targeted supporting meaningful social connections, findings highlighted the significant role of the VE and its elements on outcomes. Baker and colleagues (2021) designed and integrated three features as social scaffolds into their ICVE, which helped stimulate spontaneous and natural interactions among participants. Meanwhile, Kalantari and colleagues (2023) found that incorporating design functionalities that evoke pleasure and excitement enhances social outcomes. These findings align with research on social VR platforms, which shows the predictive role of enhanced social and spatial presence—impacted by design elements and activities of the VE—in inducing psychological benefits such as relatedness, enjoyment, and perceived social support (Barreda-Ángeles & Hartmann, 2022; van Brakel et al., 2023).

The impact of the ICVE's design on social and health outcomes further accentuates the benefit of multidisciplinary endeavours and the need for co-creation of the environment with participants. While the majority of included studies incorporated varying degrees of co-creation, three articles undertook and reported on intensive co-design processes, resulting in refined interventions and engaging experiences (Baker et al., 2021; Matsangidou et al., 2022; Shah et al., 2023). The fruitfulness of these efforts is reflected in how several design choices were modified after prototype iterations with and feedback from participants.

The role of avatars in shared environments extends beyond mere appearances; previous research found that their characteristics affect users' behaviours and attitudes in the VE accordingly, a phenomenon known as the proteus effect (Yee & Bailenson, 2007). As a factor also influential in enhancing copresence, the behavioural realism of users' representations plays an important part in developing and projecting an identity to the avatars. To achieve highly realistic gestures and expressions reflecting users' real-world movements, real-time tracking systems are essential. Two of the included studies (Baker et al., 2021; Matsangidou et al., 2022) incorporated tracking of facial expressions on top of regular head and hands tracking supported by newer HMD models. This consideration is particularly important in health programs which aim to take advantage of the social and interactive aspects of ICVEs. Furthermore, providing the opportunity to observe self-avatars in virtual mirrors is another strategy to induce a sense of agency and identity over avatars. This was embedded in three of the included studies as part of the pre-program training sessions (Baker et al., 2021; Barberia et al., 2018) and as a design element over the course of the program (Matsangidou et al., 2022).

## Limitations

Despite the insightful findings of the included studies, a mindful stance should be adopted in their appraisal. As seen in Tables 1 and 3, only four out of the ten studies included a control group in their methodology, which, rather than reflecting a lack of rigour, emphasises the preliminary stage of research in the field. Accordingly, seven out of ten studies were published in the year 2023, indicating the nascent nature of the research topic. Furthermore, on average, 27 participants were involved across the studies, ranging from three in a case study to 64 in an RCT. This varied adoption of research designs, aims, and health outcomes prevents us from reaching any conclusion on the superiority of ICVE interventions' health benefits over any other approach based on the included studies in this review. However, clear implications and impetus for further research can be drawn from the reported qualitative and quantitative findings.

## Conclusion

Upon examining the literature on the use of immersive technologies in various health applications, many reviews attempt to establish their effectiveness, albeit most report the same limitations of lacking control groups and small sample sizes. Despite the limitations, these technologies remain particularly attractive to the healthcare researcher interested in influencing and leveraging participants' behavioural responses and social interactions in simulated environments. These technologies promise to provide more realistic and ecologically valid setups without sacrificing the experimental control of a laboratory setting (Blascovich et al., 2002). This is supported by research showing similarity between participants' real-world physical and their embodied virtual behavioural reactions via avatars (e.g., Yee et al., 2007).

However, this rather unique advantage comes with the caveat of engendering the assumption of veridicality. This means that if experiences and responses within these immersive environments are sufficiently realistic and authentic to evoke behaviours in users similar to those happening in the real world, then the skills and patterns of behaviour learned within these environments are assumed to be automatically generalisable to real-world settings (Parsons, 2016). Mindlessly adopting this assumption can give rise to a technocentric perspective, overemphasising the role of technologies in solving human problems (Schmidt & Glaser, 2021). This contrasts with the user-centred approach, which advocates for contextualising and customising the affordances of new media and technologies for the specific population and outcomes of interest, while actively involving participants in the process. Indeed, there needs to be an intentional intervention design process to effectively utilise the technology to our benefit, considering the interplay of factors in the technology-participant-health network (see Figure 2). In other words, to fully realise the potential of these systems for participants, health researchers should consider 'which technologies work for whom, in which contexts, with what kinds of support, and for what kinds of tasks or objectives?' (Parsons, 2016, p. 153). Accordingly, the review of included studies has shown how researchers' decisions about the factors related to each element of the aforementioned network can influence achieving the main objectives.

In conclusion, we reviewed ten studies that used ICVEs to bring about various health benefits to participants. The combination of quantitative and qualitative data across studies demonstrated that ICVE interventions resulted in varying degrees of positive health outcomes, including well-being promotion, symptom reduction, and skill acquisition. Notably, all studies benefited from ICVEs' potential in enabling social interactions towards their diverse health aims, including four studies with older adults. Furthermore, participants reported mostly positive experiences and were generally accepting of the technology. They mentioned high usability scores, motivation, enjoyment, engagement, and heightened presence and copresence within the VEs. However, the need for more training and familiarisation time was highlighted as an area for improvement. On the one hand, no conclusive assertion can be made on the clinical effectiveness of ICVE health interventions at this stage due to the heterogeneous research designs and aims across studies. On the other hand, the role of an intentional intervention design, considering factors affecting presence and copresence within the immersive environment, as well as integrating co-creation of the program with participants, seems integral to successfully achieving the aim of 'being there together for health'.

## Acknowledgment

TZ was supported by the Australian Government Research Training Program Scholarship.

Systematic Review on Immersive Collaborative Virtual Environments in Health